# Point group symmetry of cadmium arsenide thin films determined by convergent beam electron diffraction

Honggyu Kim[1], Manik Goyal[1], Salva Salmani-Rezaie[1], Timo Schumann[1], Tyler N. Pardue[1], Jian-Min Zuo[2,3] and Susanne Stemmer[1,*]

[1]Materials Department, University of California, Santa Barbara, California 93106-5050, USA

[2]Department of Materials Science and Engineering, University of Illinois at Urbana-Champaign, Urbana, Illinois 61801, USA

[3]Frederick Seitz Materials Research Laboratory, University of Illinois at Urbana-Champaign, Urbana, Illinois 61801, USA

*Corresponding author. Email: stemmer@mrl.ucsb.edu

**ABSTRACT**

Cadmium arsenide ($Cd_3As_2$) is one of the first materials to be discovered to belong to the class of three-dimensional topological semimetals. Reported room temperature crystal structures of $Cd_3As_2$ reported differ subtly in the way the Cd vacancies are arranged within its antifluorite-derived structure, which determines if an inversion center is present and if $Cd_3As_2$ is a Dirac or Weyl semimetal. Here, we apply convergent beam electron diffraction (CBED) to determine the point group of $Cd_3As_2$ thin films grown by molecular beam epitaxy. Using CBED patterns from multiple zone axes, high-angle annular dark-field images acquired in scanning transmission electron microscopy, and Bloch wave simulations, we show that $Cd_3As_2$ belongs to the tetragonal 4/*mmm* point group, which is centrosymmetric. The results show that CBED can distinguish very subtle differences in the crystal structure of a topological semimetal, a capability that will be useful for designing materials and thin film heterostructures with topological states that depend on the presence of certain crystal symmetries.

**INTRODUCTION**

Over the past few years, three-dimensional Dirac and Weyl semimetals have emerged as an important class of topological materials [1-3]. They exhibit band crossings ("nodes") with linear energy dispersion in all three dimensions of momentum space. Dirac semimetals possess both time reversal and inversion symmetry. In contrast, one or both of these symmetries must be absent in a Weyl semimetal, with profound effects on their properties. For example, in a Weyl semimetal, nodes appear in pairs of opposite chirality because of the lack of spin degeneracy. As a result, they feature topologically protected surface Fermi arcs that connect projections of nodes of opposite chirality [1-3]. In contrast, the nodes of three-dimensional Dirac semimetals are spin degenerate and each Dirac node can be considered as two overlapping Weyl nodes of opposite chirality. The surfaces of these Dirac semimetals feature double Fermi arc-like features that are less protected and can be disconnected from the projection of the nodes [4].

Cadmium arsenide ($Cd_3As_2$) is one of the most important topological semimetals discovered to date, because it features a single pair of nodes and no other (trivial) bands at the Fermi level, besides many other attractive properties, such as a high Fermi velocity [5-11]. The room temperature crystal structure of $Cd_3As_2$ is body-centered tetragonal [12], with the four-fold rotational symmetry being essential in protecting the band crossings against the opening of a gap [5]. The structure can be described as being derived from an antifluorite structure with an ordered arrangement of vacant Cd sites, allowing for the 3:2 Cd:As stoichiometry. The Cd vacancies cause the systematic displacements of the occupied sites from their ideal antifluorite positions and a large unit cell with dimensions that are approximately doubled along the *a*-axis (12.63 Å) and quadrupled along the *c*-axis (25.43 Å), relative to the cubic antifluorite cell. Important details of the room temperature crystal structure remain, however, somewhat controversial. Specifically, $Cd_3As_2$ has been reported as belonging to either space group $I4_1cd$ [5, 12-14] or $I4_1/acd$ [15]. The differences between the two proposed structures are subtle and arise only from the Cd vacancy arrangements (and associated displacements). Furthermore, the two structures differ in a key aspect: $I4_1cd$ (point group 4*mm*) lacks an inversion center, making $Cd_3As_2$ a Weyl semimetal, as originally proposed [5], while $I4_1/acd$ (point group 4/*mmm*) is centrosymmetric and is consistent with the more common description of $Cd_3As_2$ as a three-dimensional Dirac semimetal. All of these studies used kinematical x-ray diffraction, for which determining the absence of an inversion symmetry can be challenging, due to Friedel's law [16].

Given the small differences between the proposed structures and the fact that $Cd_3As_2$ undergoes multiple structural transitions with temperature [17], it is also conceivable that materials synthesized by different methods or with slightly different stoichiometries may have different Cd vacancy orderings. Furthermore, it is important to characterize the structure of thin films of $Cd_3As_2$, which are important for potential future applications. In particular, films are susceptible to symmetry breaking by various means, such as thermal and lattice mismatch strains, which directly affect the topological states. A method that can determine subtleties in their crystal symmetries, despite the small scattering volume, is essential in establishing structure-property relationship of thin films of topological materials.

Convergent beam electron diffraction (CBED) patterns in transmission electron microscopy (TEM) arise from dynamical scattering, making Friedel's law no longer applicable [18]. CBED has long been used to determine point and space group symmetries [18-21] and is suitable for small volumes, such as thin films. Here, we use CBED to determine the point group symmetry of epitaxial, high mobility (~19,000 $cm^2$/Vs at room temperature [22]) $Cd_3As_2$ films grown by molecular beam epitaxy (MBE) on III-V substrates. We use CBED patterns along multiple zone axes to identify the epitaxial orientation relationship and the point group symmetry. We show that MBE-grown $Cd_3As_2$ films possess an inversion center, classifying them as Dirac semimetals.

**EXPERIMENTAL**

$Cd_3As_2$ films were grown by MBE on relaxed (111) GaSb buffer layers on As-terminated (111)B GaAs substrates as described elsewhere [22]. The thickness of the $Cd_3As_2$ films was 270 nm and they grew relaxed (unstrained), which is unsurprising, given the large lattice mismatch with GaSb (~3.5%). The out-of-plane film-substrate epitaxial orientation relationship can be described as $(112)_T \| (\bar{1}\bar{1}\bar{1})_C$, where the subscripts denote the tetragonal and cubic unit cells of films and substrates/buffer layers, respectively [22]. We note the similarities in the atom arrangements on these surfaces, as depicted in Fig. 1. The hexagonal arrangement of the cubic III-V growth surface suggests three in-plane orientation relationships are possible, with either $(\bar{1}02)_T$, $(01\bar{2})_T$, or $(1\bar{1}0)_T$ parallel to $\{1\bar{1}0\}_C$, possibly leading to the formation of twins. Use of miscut substrates suppresses twinning in the $Cd_3As_2$ [23]. Here, the miscut angle was 1°.

Cross-section TEM samples were prepared by focused ion beam (FIB) thinning using a FEI Helios Dual beam nanolab 650 FIB system with a final milling energy of 2 kV Ga ions. TEM samples were prepared along three different zone axes, $[\bar{2}01]_T$, $[02\bar{1}]_T$, and $[1\bar{1}0]_T$, which are parallel to $\langle 1\bar{1}0 \rangle_C$ of the III-V semiconductors. The specific indices of these three zone axes, denoted A-C in the following, were not known a priori, but determined as described below. Figure 2(a) depicts a schematic of the heterostructure with different colors (red, green, and orange), representing different planes, which define three different zones. High-angle annular dark-field imaging (HAADF) in scanning transmission electron microscopy (STEM) and CBED were conducted using a FEI Titan S/TEM ($C_s$=1.2 mm) with 300 kV and 150 kV electrons, respectively. To reduce the background intensity from thermal diffuse scattering, all CBED patterns shown here were taken at 103 K using a double-tilt liquid nitrogen cooling holder. Both zero-order Laue zone (ZOLZ) and whole patterns (WPs), which include high-order Laue zone (HOLZ) rings, were recorded. ZOLZ patterns show the rotational and mirror symmetries of the projected (two-dimensional) structure, while HOLZ rings contain information about three-dimensional symmetries. The experimental patterns were analyzed following the methodology and nomenclature by Buxton et al. [19]. Simulations of the CBED patterns were carried out using a Bloch wave algorithm as implemented by Zuo et al. [21, 24].

**RESULTS AND DISCUSSION**

Figures 2(b-d) show atomic resolution HAADF-STEM images for zones A-C. Columns containing Cd vacancies (light blue dots) can be identified by their reduced intensities and are located in alternating rows (see white arrows). Half of the Cd columns contain vacancies, consistent with our previous report [22] and the reported crystal structures for $Cd_3As_2$ at room temperature [12, 15]. The three zone axes, $[\bar{2}01]_T$, $[02\bar{1}]_T$, and $[1\bar{1}0]_T$, cannot easily be distinguished in these images, because the arrangements of Cd vacancies appear similar in projection in both reported crystal structures.

Figure 3 shows the experimental ZOLZ patterns and WPs for zones A-C. All CBED patterns are oriented such that the growth direction, $[221]_T$, points upward. A notable difference in the WPs is the radius of the first-order Laue zone (FOLZ) ring, which is larger in zone C than in zones A and B. This identifies zone C as $[1\bar{1}0]_T$, because the radius of HOLZ rings reflects the reciprocal lattice spacing (*H*) perpendicular to the electron beam direction. For tetragonal structures, $H = [a^2(u^2 + v^2) + w^2c^2]^{-1/2}$, where [$u$, $v$, $w$] are the zone axes indices and $a$ and $c$ are the lattice parameters, which is largest for $[1\bar{1}0]_T$. Using the out-of-plane orientation information, the CBED patterns from zones A and B can then be identified as $[\bar{2}01]_T$ and $[02\bar{1}]_T$, respectively.

As indicated in Fig. 3, the 000 discs ("bright-field pattern" or BP) and WP along $[\bar{2}01]_T$ and $[02\bar{1}]_T$ possess a single mirror (*m*), while those along $[1\bar{1}0]_T$ show two orthogonal mirrors and a two-fold rotational axis and thus have 2*mm* symmetry. Referring to the tables by Buxton [19], the possible CBED diffraction groups having one mirror plane in both BP and WP are *m* and $2_R m m_R$, where the symbols $2_R$ and $m_R$ refer to a two-fold horizontal axis and an inversion center, respectively. The possible diffraction groups corresponding to 2*mm* symmetry in both BP and WP are 2*mm* and $2mm1_R$, where $1_R$ refers to a horizontal mirror perpendicular to the electron beam direction. Along [*u*0*w*], materials belonging to space groups $I4_1/acd$ (point group 4/*mmm*) and $I4_1cd$ (point group 4*mm*) display diffraction groups $2_R m m_R$ and *m*, respectively. Both groups have one mirror plane in BP and WP and are thus indistinguishable in [*u*0*w*] patterns. Along $[1\bar{1}0]_T$, however, the point groups can be distinguished in the WPs: 4/*mmm* is expected to show 2*mm* symmetry in both BP and WP, while 4*mm* shows 2*mm* symmetry in BP but only *m* in WP. Here, both BP and WP along $[1\bar{1}0]_T$ show 2*mm* symmetry, indicating that the $Cd_3As_2$ films belong to the centrosymmetric 4/*mmm* point group. Additional confirmation comes from symmetries (two mirrors and two-fold rotational axis, 2*mm*) that are present in a <100> WP, which is also consistent with 4/*mmm* (see Supplementary Material [25]).

To confirm the analysis based on the Buxton tables, we also performed Bloch wave simulations of CBED patterns for both $I4_1/acd$ and $I4_1cd$ structures. As shown in Fig. 4(a), the symmetry elements of BPs and WPs seen in the calculated CBED for the three zone axes of the $I4_1/acd$ structure correspond to those observed in experimental CBED patterns in Fig. 3: $m$ in BPs and WPs for $[\bar{2}01]_T$ and $[02\bar{1}]_T$; $2mm$ in BP and WP for $[1\bar{1}0]_T$. In contrast, the $[1\bar{1}0]_T$ CBED pattern of the $I4_1cd$ structure, shown in Fig. 4(b) (bottom) shows only one mirror in WP, inconsistent with our experimental results. Thus, the simulated patterns further confirm our conclusion that thin films of $Cd_3As_2$ possess an inversion center.

The analysis above was primarily focused on distinguishing point groups $4/mmm$ and $4mm$ corresponding to the two reported crystal structures for $Cd_3As_2$. CBED allows, however, for determining the point group even without any prior knowledge. Table I summarizes the possible diffraction groups using the symmetries observed along $[\bar{2}01]_T$, $[02\bar{1}]_T$, and $[1\bar{1}0]_T$. The five possible point groups that correspond to the diffraction groups are: $mm2$, $4/mmm$, $6/mmm$, $m3$ and $m3m$. The point group can be identified by recording an additional different CBED pattern along a high-symmetry zone axis. Figure 5 shows a CBED pattern along $[001]_T$. Both BP and WP display two mirror planes and four-fold rotational symmetry ($4mm$). The possible diffraction groups are $4mm$ and $4mm1_R$. In combination with the information from the other zone axes, this leaves two possible point groups, $4/mmm$ and $m3m$. The $m3m$ point group is cubic, which is the structure adopted by $Cd_3As_2$ at high temperatures, and which has randomly distributed Cd vacancies [17, 26]. This is in contrast with the HAADF-STEM images, which show that the Cd vacancies are systematically ordered. In addition, in a cubic structure, the three zone axes (A-C) would have the same symmetries and HOLZ ring diameters, which is not seen in our experiments.

Lastly, we note that the intermediate-temperature modification of $Cd_3As_2$ (space group $P4_2/nmc$) also belongs to point group $4/mmm$ [15, 17]. Although we do not expect this phase in our films, which are grown at temperatures below the transition [22], it may conceivably persist as a metastable phase in other types of samples. The intermediate temperature phase differs substantially from the low temperature modification in the way the Cd vacancies are arranged, leading to distinct features in HAADF-STEM images, such as channels of Cd vacancies channels along $\langle 100 \rangle_T$ (see Supplementary Information [25]). In case of our films, HAADF-STEM images recorded along zone axes $[001]_T$ and $[010]_T$ are not consistent with this intermediate-temperature modification (see Supplementary Information [25]), but show excellent agreement with the I4$_1$/*acd* structure, as discussed above.

**CONCLUSIONS**

In conclusion, we have shown that MBE-grown Cd$_3$As$_2$ films are centrosymmetric and belong to point group 4/*mmm*. In combination with the Cd-vacancy ordered patterns observed in HAADF-STEM this demonstrates that the films have the same crystal structure as reported in ref. [15] for Cd$_3$As$_2$ single crystals. This shows that the structure of the films is not modified by any of the parameters of thin film growth, such as residual strain, low growth temperature etc. Thus, predictions from electronic structure theory based on the reported, centrosymmetric crystal structure, should directly apply to these films. Furthermore, the results show that CBED should be able to detect broken symmetries, such as due to epitaxial coherency strains [27] or near surfaces and interfaces. Because the nodes in these semimetals are protected by both symmetry and topology [2], such studies will provide essential information relevant for topological state engineering.

**Acknowledgments**

The authors thank Prof. Les Allen and Dr. Y.-T. Shao for helpful suggestions and discussions regarding CBED simulation and analysis. The electron microscopy experiments were supported by the U.S. Department of Energy (Grant No. DEFG02-02ER45994). Film growth experiments were supported through the Vannevar Bush Faculty Fellowship program by the U.S. Department of Defense (Grant No. N00014-16-1-2814). This work made use of the MRL Shared Experimental Facilities, which are supported by the MRSEC Program of the US National Science Foundation under Award No. DMR 1720256.

# References

[1] T. O. Wehling, A. M. Black-Schaffer, and A. V. Balatsky, Adv. Phys. **63**, 1 (2014).

[2] B.-J. Yang, and N. Nagaosa, Nat. Comm. **5**, 4898 (2014).

[3] N. P. Armitage, E. J. Mele, and A. Vishwanath, Rev. Mod. Phys. **90**, 015001 (2018).

[4] M. Kargarian, M. Randeria, and Y. M. Lu, Proc. Natl. Acad. Sci. **113**, 8648 (2016).

[5] Z. J. Wang, H. M. Weng, Q. S. Wu, X. Dai, and Z. Fang, Phys. Rev. B **88**, 125427 (2013).

[6] M. Neupane, S. Y. Xu, R. Sankar, N. Alidoust, G. Bian, C. Liu, I. Belopolski, T. R. Chang, H. T. Jeng, H. Lin, A. Bansil, F. Chou, and M. Z. Hasan, Nat. Comm. **5**, 3786 (2014).

[7] S. Jeon, B. B. Zhou, A. Gyenis, B. E. Feldman, I. Kimchi, A. C. Potter, Q. D. Gibson, R. J. Cava, A. Vishwanath, and A. Yazdani, Nat. Mater. **13**, 851 (2014).

[8] T. Liang, Q. Gibson, M. N. Ali, M. H. Liu, R. J. Cava, and N. P. Ong, Nat. Mater. **14**, 280 (2015).

[9] S. Borisenko, Q. Gibson, D. Evtushinsky, V. Zabolotnyy, B. Buchner, and R. J. Cava, Phys. Rev. Lett. **113**, 165109 (2014).

[10] Z. K. Liu, J. Jiang, B. Zhou, Z. J. Wang, Y. Zhang, H. M. Weng, D. Prabhakaran, S.-K. Mo, H. Peng, P. Dudin, T. Kim, M. Hoesch, Z. Fang, X. Dai, Z. X. Shen, D. L. Feng, Z. Hussain, and Y. L. Chen, Nat. Mater. **13**, 677 (2014).

[11] I. Crassee, R. Sankar, W.-L. Lee, A. Akrap, and M. Orlita, Phys. Rev. Mater. **2**, 120302 (2018).

[12] G. A. Steigmann, and J. Goodyear, Acta Cryst. **B24**, 1062 (1968).

[13] R. Sankar, M. Neupane, S. Y. Xu, C. J. Butler, I. Zeljkovic, I. P. Muthuselvam, F. T. Huang, S. T. Guo, S. K. Karna, M. W. Chu, W. L. Lee, M. T. Lin, R. Jayavel, V. Madhavan, M. Z. Hasan, and F. C. Chou, Sci. Rep. **5**, 12966 (2015).

[14] H. M. Yi, Z. J. Wang, C. Y. Chen, Y. G. Shi, Y. Feng, A. J. Liang, Z. J. Xie, S. L. He, J. F. He, Y. Y. Peng, X. Liu, Y. Liu, L. Zhao, G. D. Liu, X. L. Dong, J. Zhang, M. Nakatake, M. Arita, K. Shimada, H. Namatame, M. Taniguchi, Z. Y. Xu, C. T. Chen, X. Dai, Z. Fang, and X. J. Zhou, Sci. Rep. **4**, 6106 (2014).

[15] M. N. Ali, Q. Gibson, S. Jeon, B. B. Zhou, A. Yazdani, and R. J. Cava, Inorg. Chem. **53**, 4062−4067 (2014).

[16] G. Friedel, C.R. Acad. Sci. Paris **157**, 1533 (1913).

[17] A. Pietraszko, and K. Lukaszewicz, Acta Cryst. B **B 25**, 988 (1969).

[18] P. Goodman, and G. Lehmpfuhl, Acta Cryst. **A 24**, 339 (1968).

[19] B. F. Buxton, J. A. Eades, J. W. Steeds, and G. M. Rackham, Philos. Trans. R. Soc. A **281**, 171 (1976).

[20] M. Tanaka, R. Saito, and H. Sekii, Acta Crystallogr. Sect. A **39**, 357 (1983).

[21] J. C. H. Spence, and J. M. Zuo, *Electron Microdiffraction* (Plenum Press, New York, 1992).

[22] T. Schumann, M. Goyal, H. Kim, and S. Stemmer, APL Mater. **4**, 126110 (2016).

[23] M. Goyal, L. Galletti, S. Salmani-Rezaie, T. Schumann, D. A. Kealhofer, and S. Stemmer, APL Mater. **6**, 026105 (2018).

[24] J. M. Zuo, and J. C. Mabon, Microscopy and Microanalysis **10**, 1000 (2004).

[25]   See Supplemental Material at [link by publisher] for CBED along $[010]_T$, HAADF-STEM images along $[001]_T$ and $[010]_T$, respectively, and a discussion of the intermediate temperature structure.

[26]   A. Pietrasz, and K. Lukaszew, Phys. Stat. Sol. A **18**, 723 (1973).

[27]   M. Goyal, H. Kim, T. Schumann, L. Galletti, A. A. Burkov, and S. Stemmer, Phys. Rev. Mater. (in press) (2019).

**Table I.** Diffraction groups and point groups in zone axis CBED patterns

| Zone axis | BP | WP | Possible diffraction groups | Possible point groups |
|---|---|---|---|---|
| $[\bar{2}01]_T$, | $m$ | $m$ | $m, 2_Rmm_R$ | $m, mm2, 4mm, \bar{4}2m, 3m, \bar{6}, 6mm, \bar{6}m2, \bar{4}3m, 2/m, mmm, 4/m, 4/mmm, \bar{3}m, 6/m, 6/mmm, m3, m3m$ |
| $[02\bar{1}]_T$ | $m$ | $m$ | $m, 2_Rmm_R$ | $m, mm2, 4mm, \bar{4}2m, 3m, \bar{6}, 6mm, \bar{6}m2, \bar{4}3m, 2/m, mmm, 4/m, 4/mmm, \bar{3}m, 6/m, 6/mmm, m3, m3m$ |
| $[1\bar{1}0]_T$ | $2mm$ | $2mm$ | $2mm, 2mm1_R$ | $mm2, \bar{6}m2, mmm, 4/mmm, 6/mmm, m3, m3m$ |
| $[001]_T$ | $4mm$ | $4mm$ | $4mm, 4mm1_R$ | $4mm, 4/mmm, m3m$ |

**Figure Captions**

**Figure 1:** Schematic (top) of the atom arrangements near Cd$_3$As$_2$ (112)$_T$ and GaSb ($\bar{1}\bar{1}\bar{1}$)$_C$ surfaces, respectively. The black dashed lines highlight similar hexagonal arrangements of group V elements in these planes. The possible in-plane epitaxial orientation relationships are shown in the bottom row. The $\{1\bar{1}0\}_C$ planes of GaSb (shown on the right) can be parallel to either one of three planes of Cd$_3$As$_2$: $(\bar{1}02)_T$, $(01\bar{2})_T$, or $(1\bar{1}0)_T$. These are shown on the left, along with their plane normal, which define the zone axes used in this study.

**Figure 2:** (a) Schematic of the heterostructure. (b-d) HAADF-STEM images of the Cd$_3$As$_2$ film along three different zone axes, corresponding to either $[\bar{2}01]_T$, $[02\bar{1}]_T$, and $[1\bar{1}0]_T$. The scale bar corresponds to 1 nm. The HAADF-STEM images in (b-d) display similar atomic arrangements, in particular, the Cd atomic columns containing Cd vacancies (blue circles). Note the color coding of image frames in (b-d) and planes in (a), respectively.

**Figure 3:** (a) Experimental CBED patterns along the three zone axes A-C. Shown are ZOLZ patterns (top) and WPs (bottom) with HOLZ rings. The growth direction, $[221]_T$, points upward. Note that different false color scales are used for ZOLZ patterns and WPs to better show the symmetries. Parts of FOLZ in the WP along $[1\bar{1}0]_T$ are magnified in (b), to better show the presence of two orthogonal mirrors.

**Figure 4:** Simulated CBED patterns of Cd$_3$As$_2$ for the two different crystals structures reported in the literature [12, 15], which correspond to two different space groups: I4$_1$/*acd*

(centrosymmetric, left) and I4$_1$cd (non-centrosymmetric, right). Note that the mirror planes indicated in the ZOLZ patterns indicate the symmetry elements present in 000 discs only. Parts of the FOLZ in the CBED pattern along $[1\bar{1}0]_T$ for the I4$_1$cd structure are magnified in (c), which confirms that there is only a single mirror plane in the WP.

**Figure 5.** Experimental CBED patterns along $[001]_T$, showing two mirror planes and four-fold rotational symmetry in both BP and WP. SOLZ denotes the second-order Laue zone.

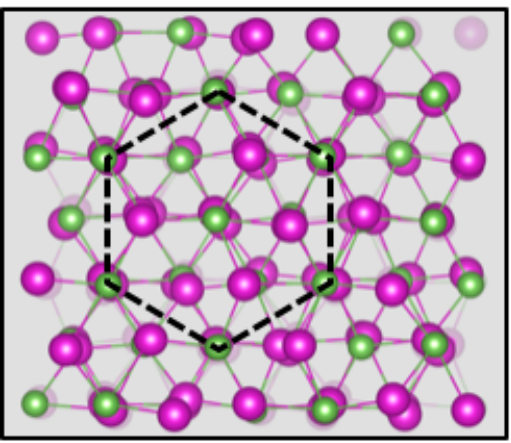
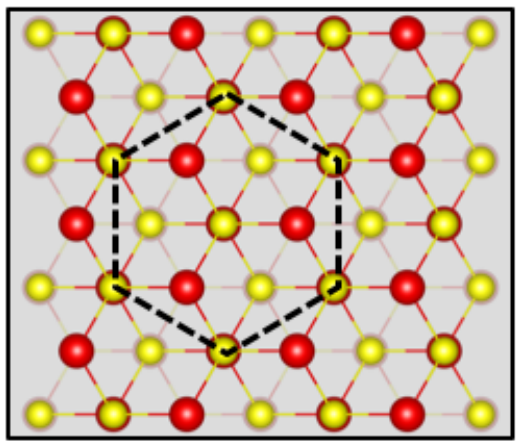
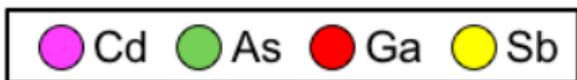
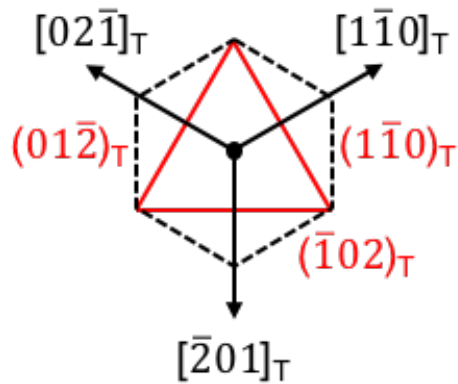
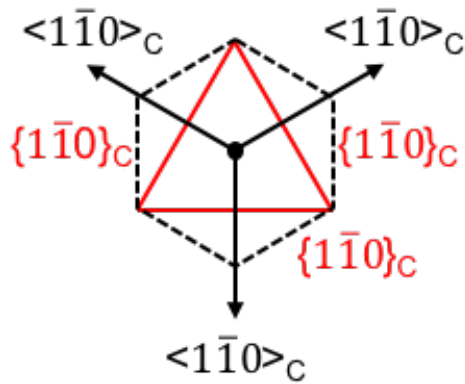

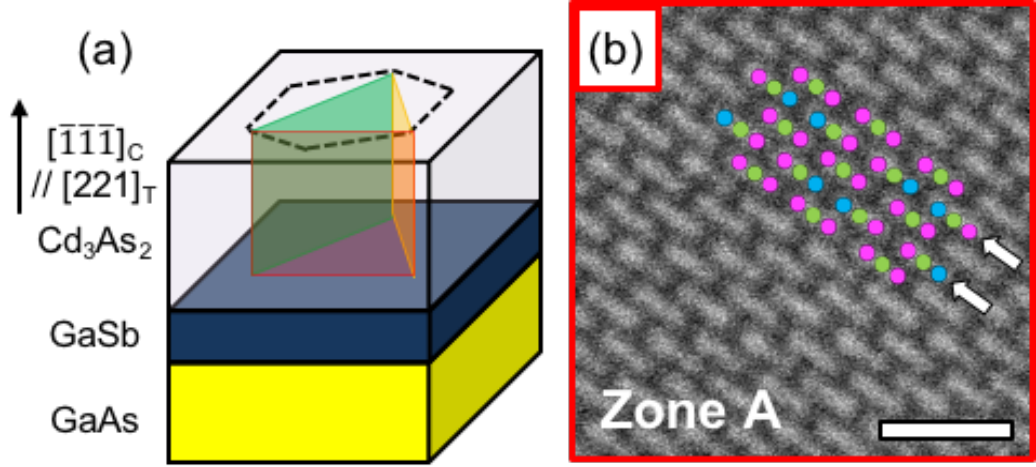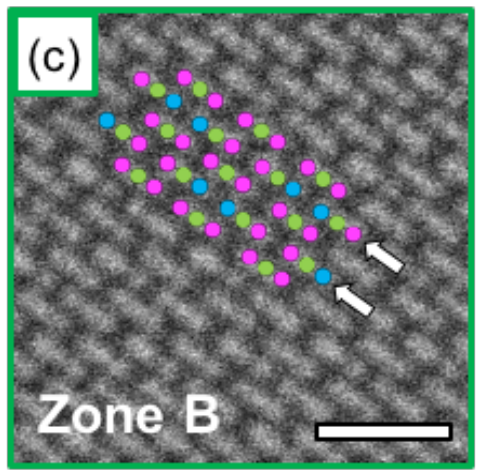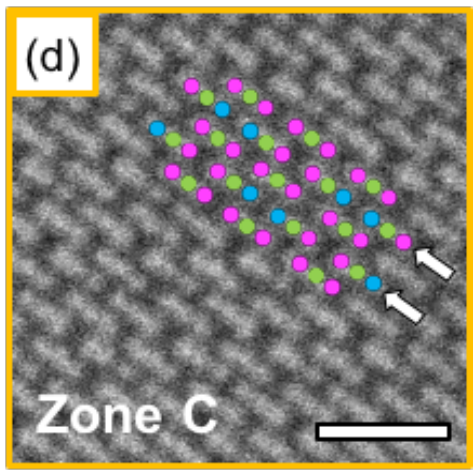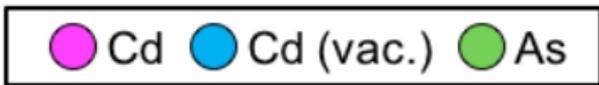

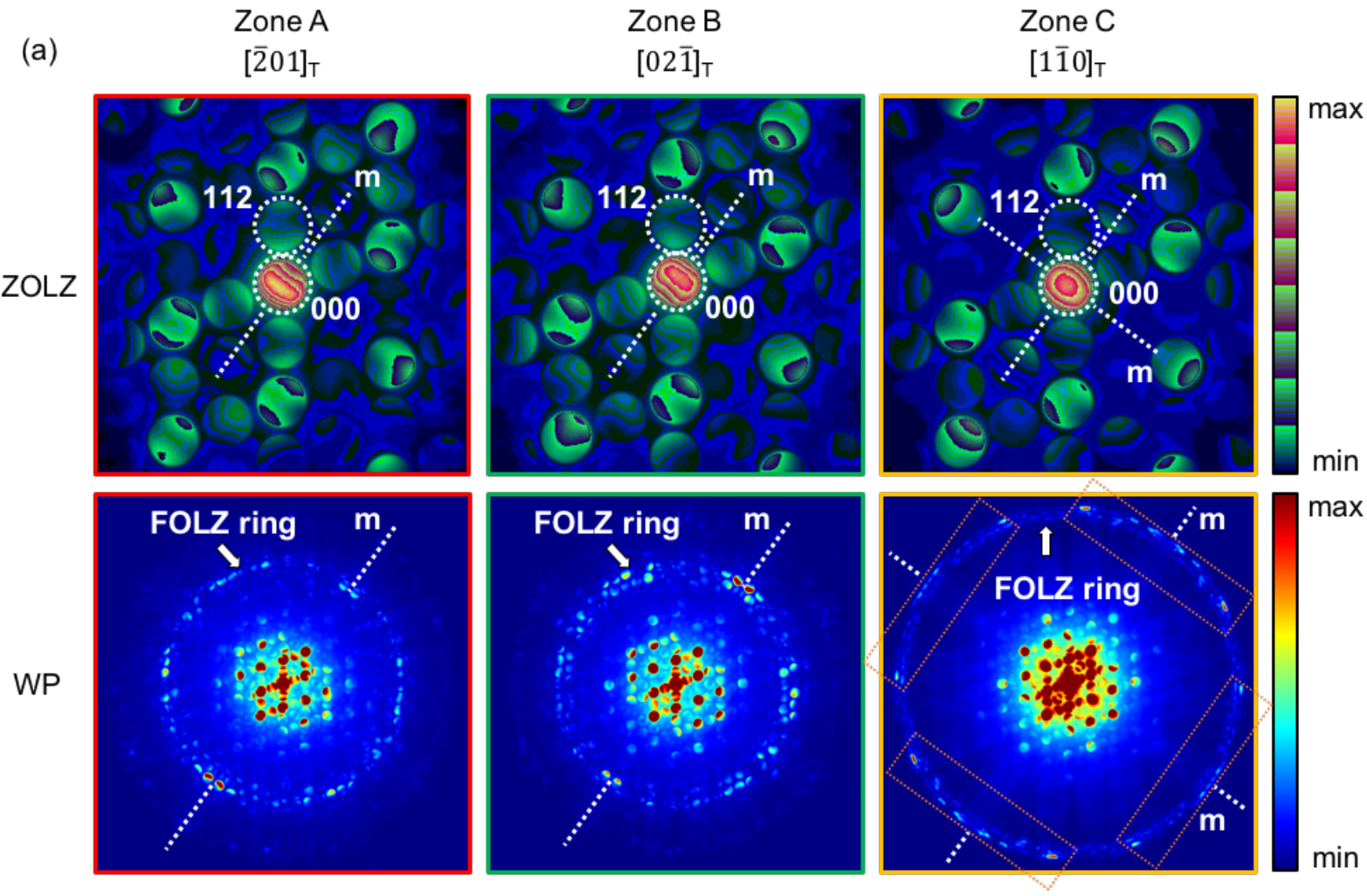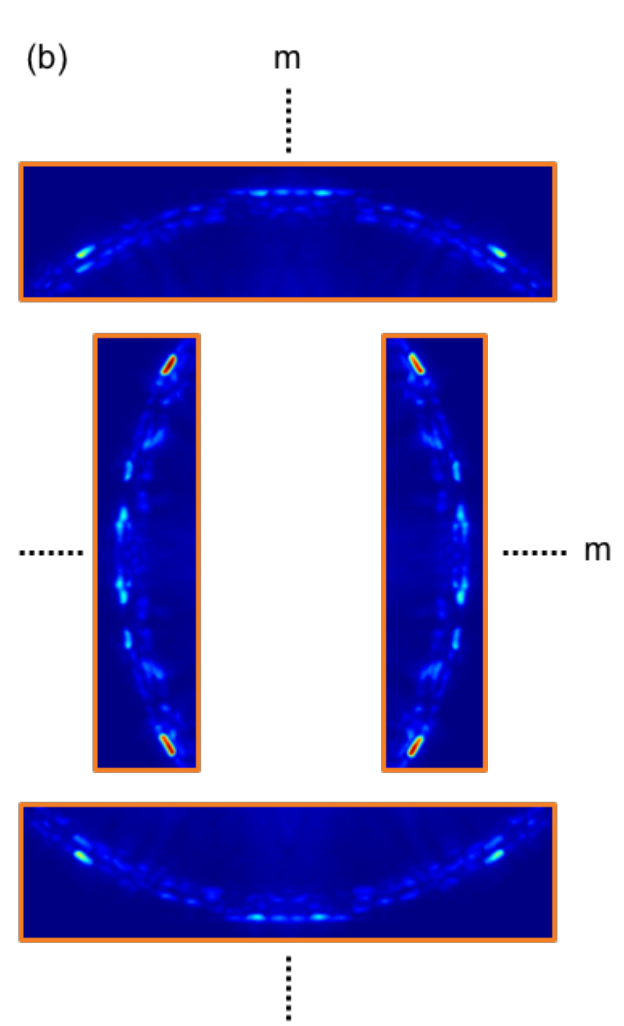

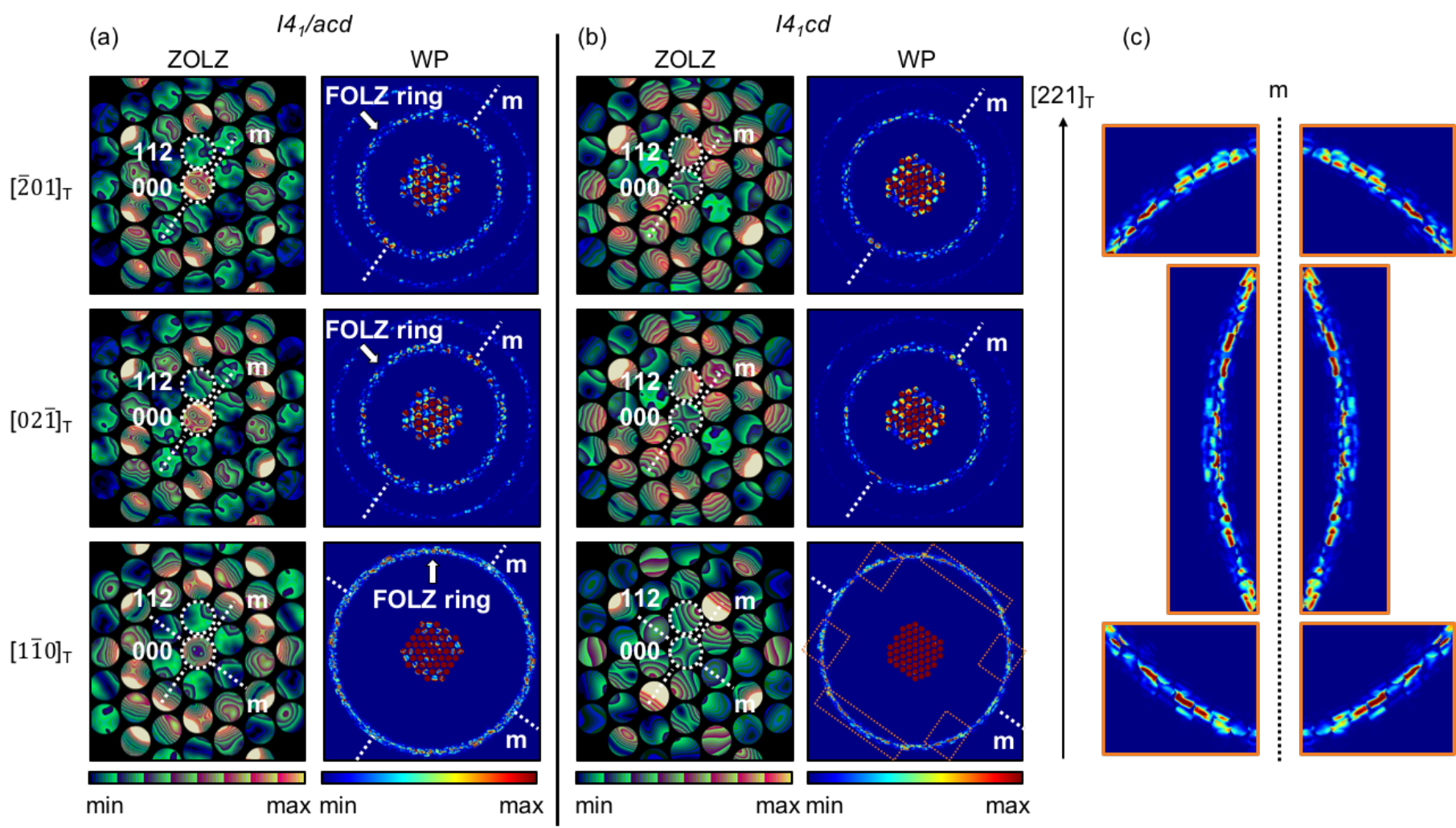

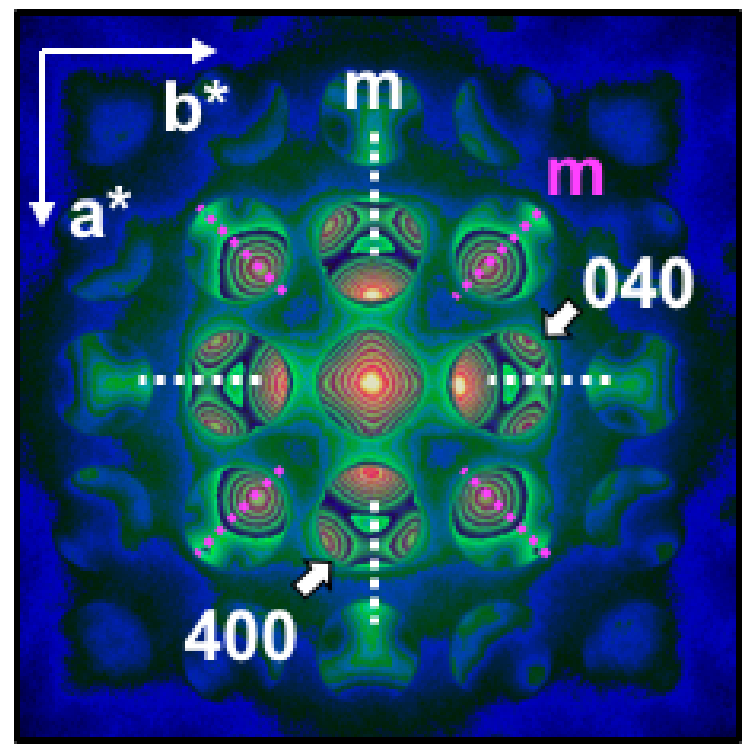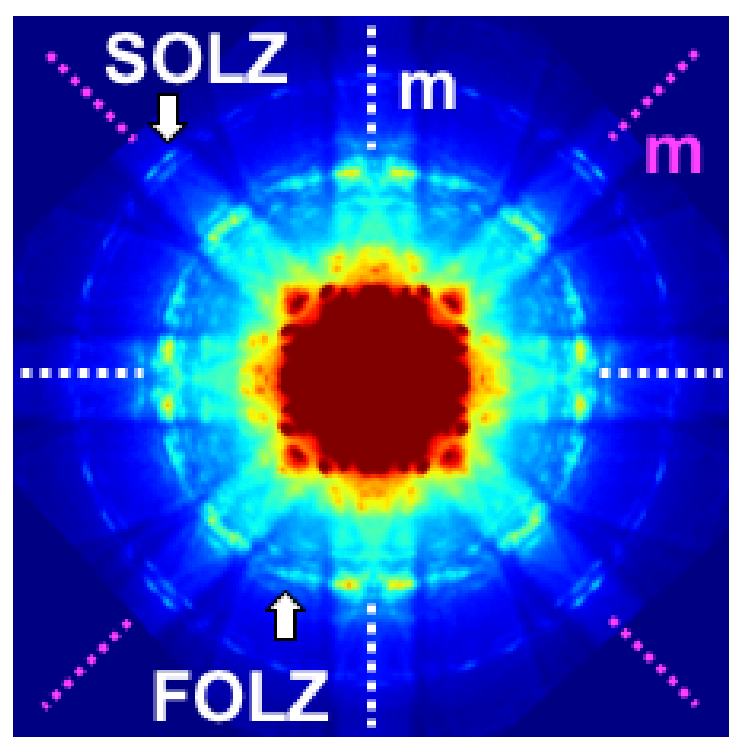